\documentclass[aps,showpacs,onecolumn,floatfix,amsmath,amssymb,nofootinbib]{revtex4-1}

\usepackage{amssymb}
\usepackage{amsmath}
\usepackage{epsfig}
\usepackage{graphicx}
\usepackage{color}
\usepackage{latexsym}
\usepackage{bm,latexsym}
\usepackage{mathrsfs}
\usepackage{float}
\usepackage{pbox}

\begin{document}

\title{Black hole-wormhole transition in (2+1)-dimensional Einstein-anti-de Sitter gravity coupled to nonlinear electrodynamics}

\author{Pedro Ca\~nate$^{1,}$$^{2,}$$^{3}$}
\email{pcanate@fis.cinvestav.mx, pcannate@gmail.com }

\author{Nora Breton$^{1}$}
\email{nora@fis.cinvestav.mx}

\affiliation{$^{1}$Departamento de F\'isica, Centro de Investigaci\'on y de Estudios Avanzados
del I.P.N.,\\ Apdo. 14-740, Mexico City, Mexico.\\
$^{2}$Facultad de Ciencias B\'asicas, Universidad Tecnol\'ogica de Bol\'ivar, Campus Tecnol\'ogico Km 1. V\'ia Turbaco, Cartagena 130001, Colombia\\ 
$^{3}$Instituto de Ciencias Nucleares,\\ 
Universidad Nacional Aut\'onoma de M\'exico, 
Apdo.  70- 543, Mexico City 04510, Mexico. \\
}

\begin{abstract} 
In this paper we present two results in $(2+1)$ gravity coupled to nonlinear electrodynamics.
First it is determined the general form of the electromagnetic field  tensor in $(2+1)$ gravity coupled to nonlinear electrodynamics in stationary cyclic spacetimes. Secondly, 
it is determined a family of exact solutions in $(2+1)$ gravity sourced by a nonlinear electromagnetic field.
The solutions are characterized by five parameters: mass $M$, angular momentum $J$, cosmological constant $\Lambda$ and two electromagnetic charges, $q_{\alpha}$ and $q_{\beta}$. Remarkably, the solution can be interpreted as a traversable wormhole, provided the fulfillment of certain inequalities by the characteristic parameters; fine tunning of the cosmological constant leads to an extreme black hole, whereas by switching off  one of the electromagnetic charges, we obtain the Ba\~nados-Teitelboim-Zanelli (BTZ) black hole.
\end{abstract}


\maketitle

\section{Introduction}

Gravity in $(2+1)$ dimensions has attracted a lot of attention due to the deep connection to a Yang-Mills theory with the Chern-Simons term \cite{Witten2007}, \cite{Achucarro1986}.
Moreover, taking advantage of simplifications due to the dimensional reduction, three dimensional Einstein theory of gravity has turned out a good model from which extract relevant insights regarding the quantum nature of gravity \cite{Carlip1998}.  In three spacetime dimensions, general relativity becomes a topological field theory with only a few nonpropagating degrees of freedom. Therefore it was a surprise that the $(2+1)-$dimensional BTZ black hole can have an arbitrarily high entropy \cite{Sachs2011}. Additionally, in string theory, there are near extremal black holes whose entropy can be calculated, and almost all of them have a near-horizon geometry containing the BTZ solution  \cite{BTZ1992}, \cite{BTZ1993}. Moreover, the thermodynamical properties of the BTZ black hole can be determined by several techniques, giving coincident results, these facts suggesting that the study of the BTZ thermodynamics could render some clues on the statistical mechanics of black holes \cite{Carlip2005}.
 
On the other hand nonlinear electrodynamics (NLED) has attracted interest lately for a number of reasons.  The need of NLED arises in the presence of  very strong electromagnetic fields (of the order of $10^{18}$ Volt/m),  because nonlinear effects occur that are  not described by the Maxwell electrodynamics. Nonlinear electrodynamics consists on theories derived from Lagrangians that depend  arbitarily on the two electromagnetic invariants, $F= 2(E^2-B^2)$ and $G= E \cdot B$, i.e. $L(F,G)$.  The ways in which $L(F,G)$ may be chosen are many, but there are two of them that outstand: the Euler-Heisenberg theory  \cite{Heisenberg_36},  derived from quantum electrodynamical assumptions,  that takes into account some nonlinear features like the interaction of light by light.  And the Born-Infeld theory  \cite{BI}, \cite{Pleban}, proposed originally with the aim of avoiding the singularity in the electric field and the self-energy due to a point charge, it is a  classical effective theory that describes nonlinear features arising in the presence of very strong electromagnetic fields,  where Maxwell linear superposition principle is not valid anymore. 
Interesting solutions have been derived from the Einstein gravity coupled to NLED, like,   black holes in which the curvature invariants, and electric field are regular everywhere (see for instance  \cite{Cataldo2000}); non-stationary wormholes sustained with nonlinear electromagnetic fields (see for instance  \cite{Bronnikov2017}, \cite{LoboArella2006}, \cite{Hendi}), among others. 

%%%%%%%%%%%%%%%%%%%%%%%%%%%%%%%%%%%%%%%%%%%%%%%%%%%%%%%%%%%%%%%%%%%%
Specifically, by using geometric units $G = c = \hbar = 1$, the action of the (2+1) Einstein theory (with cosmological constant) coupled to NLED with caracteristic function $L(F)$ is given by,

\begin{equation}\label{actionf}
S[g_{ab},A_{a}] = \int d^{3}x \sqrt{-g} \left( \frac{1}{16\pi}(R - 2\Lambda) + L(F)  \right).
\end{equation} 
where $R$ is the Ricci scalar and $\Lambda$ is the cosmological constant or de Sitter parameter. Varying this action with respect to gravitational field gives the Einstein equations,

\begin{equation}\label{EinsteinEqs}
G_{ab} + \Lambda g_{ab} = 8\pi E_{ab},
\end{equation}
with,

 \begin{equation}
4\pi E_{ab} = g_{ab}L(F) - f_{ac}f_{b}{}^{c}L_{F}, \label{Eab}
\end{equation}
where $E_{ab}$ is the electromagnetic energy-momentum tensor and $L_{F}$ stands for the derivative of $L(F)$ with respect to $F=\frac{1}{4}f_{ab}f^{ab}$. 
The variation with respect to the electromagnetic potential $A_{a}$ entering in $f_{ab} = 2\partial_{[a} A_{b]}$, yields the electromagnetic field equations,

\begin{equation}
\nabla_{a}(L_{F}f^{ab}) = 0.
\label{emEqs}
\end{equation}

In this paper we determine the general form of the electromagnetic field tensor $f_{ab}$ for the nonlinear electromagnetic theory derived from a Lagrangian depending on the invariant $F$ in a completely general form, $L(F)$,  in a stationary cyclic spacetime, in $(2+1)$ gravity;  this result is presented in the form of a theorem that generalizes the corresponding one for Maxwell theory, previously derived  by Ay\'on, Cataldo and Garc\'ia, (see for details \cite{Ayon,AGarcia2009,Garcia2017}). It turns out that $f_{ab}$  is characterized by three parameters, $a, b$ and $c$ that define two disjoint branches, $c \ne 0$ and $c=0$.

Then, considering the branch with $c \ne 0$, we derive a family of solutions characterized by five parameters, namely, mass $M$, angular momentum $J$, cosmological constant $\Lambda$ and two electromagnetic charges, $q_{\alpha}$ and $q_{\beta}$. The solution is sourced by a particular nonlinear Lagrangian, $L(F)= \sqrt{-sF},$   these kind of lagrangians have been called Einstein-power-Maxwell theories  \cite{Hassaine2008,Gurtug2012}. We verify that the solution can be interpreted as a wormhole (WH), provided  the characteristic parameters  fulfill some restrictions.  The uncharged limit ($q_{\alpha}=0$) of the WH leads to the BTZ black hole.
For a very particular value of the cosmological constant, $\Lambda_0$,  the solution becomes an extreme black hole; it turns out that  the effect of  $\Lambda_0$ cancels out the electromagnetic field, and the electromagnetic invariant vanishes, being then a solution in AdS background.  
If $\Lambda_0$ is incremented by a small positive quantity, $\Lambda_0 + \delta^2$, the extreme black hole metric transforms into a wormhole one. A possible interpretation of this fact is that the extreme  black hole constitutes an unstable structure,  that when  slightly  perturbed, by  changing the value of $\Lambda_0$ to $\Lambda_0 + \delta^2$ with ($\delta^2 <|\Lambda_0|$), it makes the black hole to decay into a wormhole with arbitrary cosmological constant;  some arguments are given regarding the variation of the de Sitter parameter. Finally, we address  the limit $\Lambda=0$, examining the effect on the curvature of the spacetime due only to the NLED.

The paper is organized as follows: In the next section the general form of the electromagnetic field  tensor $f_{ab}$ in $(2+1)$ gravity coupled to nonlinear electrodynamics in stationary cyclic spacetimes is determined with the corresponding proof. In section \ref{Sec2} the field equations for the stationary cyclic spacetime  with cosmological constant are posed restricting ourselves  to  the  $c \ne 0$ branch  of the electromagnetic field.  In Section \ref{Sec3}  the corresponding solution is derived,  determining 
a family of exact solutions in $(2+1)$ gravity sourced by a nonlinear electromagnetic field characterized by the Lagrangian $L(F)= \sqrt{-sF}$, where $s$ is the parameter of the theory. In  subsections  of \ref{Sec3} several aspects and limits of the derived solution are explored: In \ref{btzBH}  the uncharged limit that leads to the BTZ solution is presented. 
 In \ref{newextremeBH}  we analyze the black hole (BH) obtained when the de Sitter parameter $\Lambda$ is fixed with  the precise value that cancels out the electromagnetic field, giving $F=0$; this black hole presents a unique horizon, zero surface gravity, being then an extreme black hole (the extreme BTZ black hole).
In \ref{WHfNLED} the conditions on the parameters are established in order that the %four
five-parameter family of solutions sourced by the NLED  be interpreted as a  WH.  The transition BH-WH is argumented in \ref{transitionsBHWH}.   Whereas in \ref{limL0} the  limit $\Lambda =0$ is analyzed. Conclusions are presented in section \ref{Concl}. 
%%%%%%%%%%%%%%%%%%%%%%%%%%%%%%%%%%%%%%%%%%%%%%%%%%%%%%%%

\section{Electromagnetic field for stationary and cyclic
symmetric $(2+1)$ spacetimes in General Relativity coupled to Nonlinear Electrodynamics: a theorem}\label{Sec1} 

%%%%%%%%%%%%%%%%%%%%%%%%%%%%%%%%%%%%%%%%%%%%%%%
%\subsection{Electromagnetic field for stationary and cyclicsymmetric $(2+1)$ spacetime}\label{theorem} 

We start by mentioning some general aspects about the stationary cyclic symmetric spacetimes and the behavior of electromagnetic fields over these spacetimes. Let us consider a stationary cyclic symmetric spacetime with signature $(-,+,+)$. By the stationary symmetric, meaning that  the spacetime is endowed with $\boldsymbol{k} = \partial_{t}$, so that $\boldsymbol{k} \cdot  \boldsymbol{k} < 0$  (i.e., $ \boldsymbol{k}$ is a timelike vector), and such that $\mathcal{L}_{ \boldsymbol{k}} \boldsymbol{g} = 0$. Whereas, the cyclic symmetry  implies that  the spacetime is endowed with $\boldsymbol{m} = \partial_{\phi}$, so that  $\boldsymbol{m} \cdot \boldsymbol{m} > 0$  ($ \boldsymbol{m}$ is a spacelike vector) and such that $\mathcal{L}_{ \boldsymbol{m}} \boldsymbol{g} = 0$,  with closed integral curves from $0$ to $2\pi$, which in turn commute $[ \boldsymbol{k},  \boldsymbol{m}] = 0$. Therefore, the Killing vector fields $\boldsymbol{k}$ and $\boldsymbol{m}$ generate the group $SO(2)\times \mathbb{R}$. The electromagnetic field, described by the antisymmetric tensor field given by, $\boldsymbol{f} = \frac{1}{2}f_{\mu\nu}dx^{\mu} \wedge dx^{\nu}$, is assumed to be stationary and cyclic symmetric, i.e., $\mathcal{L}_{ \boldsymbol{k} }  \boldsymbol{f} = 0 = \mathcal{L}_{ \boldsymbol{m} } \boldsymbol{f}$.
It should be pointed out that, in contrast to the general $(3+1)-$dimensional stationary cyclic symmetric spacetime case, any $(2+1)-$dimensional stationary cyclic symmetric spacetime is necessarily circular, i.e., the circularity
conditions,
\begin{equation}
 \boldsymbol{m} \wedge  \boldsymbol{k} \wedge d  \boldsymbol{k} = 0 =  \boldsymbol{k} \wedge  \boldsymbol{m} \wedge d  \boldsymbol{m},     
\end{equation}
are identically fulfilled because of their 4-form character, and in consequence there exists the discrete symmetry when simultaneously $t \rightarrow -t$ and $\phi \rightarrow -\phi$. Hence, for these kinds of spacetimes, one may find a coordinate system such that the metric tensor components $\boldsymbol{g}(\boldsymbol{k}, \boldsymbol{\partial_{r}}) = 0$ and $\boldsymbol{g}( \boldsymbol{m}, \boldsymbol{\partial_{r}}) = 0$, where the coordinate direction $\partial_{r}$ is orthogonal to the surface spanned by, $\boldsymbol{k} \wedge  \boldsymbol{m}$. Usually in $(2+1)$ gravity the coordinate system $\{t, \phi, r \}$  is introduced. Therefore, in $(2+1)-$dimensional gravity any stationary cyclic symmetric metric ($\boldsymbol{g} = ds^{2}$) can be written as,

\begin{equation}\label{StCyc}
ds^{2} = g_{tt}dt^{2} + 2g_{t\phi}dt d\phi + g_{\phi \phi}d \phi^{2} + g_{rr}dr^{2}. 
\end{equation}
Now, the main goal of this section is to demonstrate the following theorem.

{\it\textbf{Theorem 1.}} The general form of stationary cyclic symmetric electromagnetic fields in $(2+1)$ dimensions in general relativity coupled to nonlinear electrodynamics $L(F)$, is given by,

\begin{equation}\label{Theorem1}
_{\ast} \boldsymbol{f} = \frac{g_{rr}c}{\sqrt{-g}}dr +  \frac{a}{3L_{F}}dt + \frac{b}{3L_{F}}d\phi,
\end{equation} 
where $a$, $b$ and $c$ are constant, while $\ast$ denotes the Hodge star operation (or Hodge dual) with respect to $g_{ab}$. The constants $a$, $b$ and $c$, are subjected, by virtue of the Ricci circularity conditions, to the following equations, 

\begin{equation}\label{Theorem1a}
ac = 0 = bc, 
\end{equation}
which give rise to two disjoint branches, the first being 

\begin{equation}\label{Theorem1b}
c  \neq 0,\quad a = 0 =b \quad \Rightarrow \quad _{\ast} \boldsymbol{f} = \frac{g_{rr} c}{\sqrt{-g}}dr,  
\end{equation}
while the second branch is,

\begin{equation}\label{Theorem1c}
c = 0 \quad \Rightarrow \quad _{\ast} \boldsymbol{f} = \frac{a}{3L_{F}}dr + \frac{b}{3L_{F}}d\phi,
\end{equation}
with its own sub-classes; ($a \ne  0$ with $b=0$) or ($b \ne 0$ with $a=0$).

{\it Proof.} To establish that the field $_{\ast}\boldsymbol{f}$  should have the form given by Eq. (\ref{Theorem1}) one uses the source–free NLED equations.\\
Let us consider the NLED Eqs. (\ref{emEqs}), in the stationary cyclic $(2+1)-$dimensional spacetime   (\ref{StCyc}. We shall make use of  the fact that the stationary and circularly symmetric character of the space-time and of the electromagnetic field;  $\mathcal{L}_{ \boldsymbol{m}} \boldsymbol{g} = 0 = \mathcal{L}_{ \boldsymbol{k}} \boldsymbol{g}$ and $\mathcal{L}_{ \boldsymbol{m}} \boldsymbol{f} = 0 = \mathcal{L}_{ \boldsymbol{k}} \boldsymbol{f}$,  demands that the metric coefficients and the electromagnetic field are independent of $t$ and $\phi$, therefore the functions $f^{\alpha\beta}$ and $g_{\alpha\beta}$ depend only on the radial coordinate; $r$. Thus, one obtains, 

\begin{equation}
\nabla_{\alpha}(L_{F}f^{\alpha\beta}) = \frac{1}{ \sqrt{-g} } \partial_{\alpha}(\sqrt{-g} L_{F}f^{\alpha\beta}) = 0   \quad \Rightarrow  \quad 
 \partial_{r}(\sqrt{-g} L_{F}f^{r t}) = 0,  \quad  \partial_{r}(\sqrt{-g} L_{F}f^{r \phi}) = 0. 
\end{equation}
Integrating the nonvanishing components, $f^{rt}$ and $f^{r \phi}$ we obtain, 

\begin{equation}\label{frtfrp}
f^{rt} = -\frac{b}{\sqrt{-g} L_{F}}, \quad f^{r\phi} = \frac{a}{\sqrt{-g} L_{F}}. 
\end{equation}
%%%
On  the other hand the Bianchi identities\footnote{ $f_{\alpha\beta} = \partial_{\alpha}A_{\beta} - \partial_{\beta}A_{\alpha}  \quad \Rightarrow  \quad \nabla_{[\nu}f_{\alpha\beta]} = 0  \quad \Rightarrow  \quad \nabla_{\beta}( _{\ast} \boldsymbol{f})^{\beta} = 0.$ And vice versa, i.e., $\nabla_{\beta}( _{\ast} \boldsymbol{f})^{\beta} = 0  \quad \Rightarrow  \quad \nabla_{[\nu}f_{\alpha\beta]} = 0$.\\
Therefore, $\nabla_{\![\nu}f_{\alpha\beta]} = 0  \quad \Leftrightarrow  \quad \nabla_{\beta}( _{\ast} \boldsymbol{f})^{\beta} = 0.$ } yield $\nabla_{\beta}( _{\ast} \boldsymbol{f})^{\beta} = 0,$  

\begin{equation}
\nabla_{\beta}( _{\ast} \boldsymbol{f} )^{\beta} = \frac{1}{ \sqrt{-g} } \partial_{\beta}(\sqrt{-g} ( _{\ast} \boldsymbol{f})^{\beta}) = 0.    
\end{equation}
Then, by expanding the previous equation, one finds, 

\begin{equation}\label{sfr}
 \partial_{r}(\sqrt{-g} ( _{\ast} \boldsymbol{f})^{r} ) = 0 \quad \Rightarrow \quad  ( _{\ast} \boldsymbol{f})^{r} = \frac{c}{\sqrt{-g}}.    
\end{equation}
On the another hand, the $n$-component of $_{\ast} \boldsymbol{f}$ in general can be written as, 

\begin{equation}\label{Theorem1u}
(_{\ast} \boldsymbol{f})_{n} = \frac{\sqrt{-g}}{3}  \left( f^{tr} \delta^{\phi}_{n}  + f^{r\phi} \delta^{t}_{n}  + f^{\phi t} \delta^{r}_{n}   \right), \textup{ with } n: t, r, \phi.
\end{equation}
which imply that  $(_{\ast} \boldsymbol{f})_{r} = \frac{\sqrt{-g}}{3}f^{\phi t}.$ Hence, by using (\ref{sfr}) we can find $f^{\phi t}$, this is;  

\begin{equation}\label{fpt}
f^{\phi t} = \frac{3g_{rr}c}{(\sqrt{-g})^{2}}.
\end{equation}
In this way,  gathering  Eqs. (\ref{frtfrp}) and (\ref{fpt}), the nonlinear electromagnetic field contravariant tensor is given by, 

\begin{equation}
f^{\alpha\beta} = \frac{1}{\sqrt{-g}}  \left[  \begin{array}{ccc}
0 & \frac{b}{L_{F}} & -\frac{3g_{rr}c}{\sqrt{-g}} \\
-\frac{b}{L_{F}} & 0 & \frac{a}{L_{F}} \\
\frac{3g_{rr}c}{\sqrt{-g}} & -\frac{a}{L_{F}} & 0
\end{array} \right] 
\end{equation}
Or its dual $_{\ast} \boldsymbol{f}$, as,
\begin{equation}\label{Theorem1Show}
_{\ast} \boldsymbol{f} = \frac{g_{rr} c}{\sqrt{-g}}dr +  \frac{a}{3L_{F}}dt + \frac{b}{3L_{F}} d\phi.
\end{equation}
The previous Theorem generalizes to NLED with  Lagrangian $L(F)$ the result by Garcia,  Eq. (3.2) in  \cite{AGarcia2009}, concerning the Maxwell field in the same spacetimes, with $L(F)=F$.  

Additionally, the vanishing conditions $ac=0=bc$ straightforwardly arise from the Ricci circularity conditions.
Consider the Ricci tensor $ \boldsymbol{R} = R_{ab}dx^{a}\otimes dx^{b}$, then the Ricci circularity conditions are;

\begin{equation}
 \boldsymbol{m} \wedge ( \boldsymbol{k} \wedge  \boldsymbol{R}( \boldsymbol{k})) = 0 =  \boldsymbol{k} \wedge ( \boldsymbol{m} \wedge  \boldsymbol{R}( \boldsymbol{m})),     
\end{equation}
or in terms of the interior product,

\begin{equation}
i_{ \boldsymbol{m}} (i_{\boldsymbol{k}}\!\!\!\!\!\quad _{\ast}\!\boldsymbol{R}( \boldsymbol{k})) = 0 = i_{ \boldsymbol{k}} (i_{ \boldsymbol{m}}\!\!\!\!\!\quad _{\ast}\!\boldsymbol{R}( \boldsymbol{m})).     
\end{equation}

On another side, by using $(2+1)-$dimensional Einstein field equations (with cosmological constant) coupled to nonlinear electrodynamics, the equation; 
$i_{ \boldsymbol{k}} (i_{ \boldsymbol{m}}\!\!\!\!\!\quad _{\ast}\!\boldsymbol{R}( \boldsymbol{m})) = 0$ can be written as; 

\begin{widetext}
\begin{equation}\label{Rcircularityconditions}
i_{ \boldsymbol{k}} (i_{ \boldsymbol{m}}\!\!\!\!\!\quad _{\ast}\!\boldsymbol{R}( \boldsymbol{m})) = 
i_{ \boldsymbol{k}} \left\{ i_{ \boldsymbol{m}} \!\!\!\!\!\quad _{\ast}\!\left[ \frac{1}{2} \left(  R_{\alpha}{}^{\alpha} - 2\Lambda   \right) g_{\phi \beta} dx^{\beta} + 2 \left( - L_{F}f_{\phi \alpha}
f_{\beta}{}^{\alpha} + L g_{\phi \beta}   \right) dx^{\beta}     \right] \right\} = 0.   
\end{equation}
\end{widetext}

By symmetries $ i_{ \boldsymbol{k}}(i_{ \boldsymbol{m}}\!\!\!\!\!\quad _{\ast}\!( g_{\phi \beta} dx^{\beta} ) ) = 0$, then (\ref{Rcircularityconditions}) implies that, 
$f_{\phi \alpha}f_{r}{}^{\alpha} = 0.$ Now, by using $f_{t\phi} = 3c,$ $f_{r}{}^{t} = {g_{rr}b}/({L_{F}\sqrt{-g} })$, 
$f_{r}{}^{r} = f_{\phi \phi}  = 0$ it is found $bc=0$.    

Likewise, starting with $i_{ \boldsymbol{m}} (i_{ \boldsymbol{k} } \!\!\!\!\!\quad _{\ast}\!\boldsymbol{R} ( \boldsymbol{k})) = 0$ one gets $ac=0$. Therefore, we have arrived to the vanishing conditions for the constants, Eq. (\ref{Theorem1a}), which give rise to the branches (\ref{Theorem1b}) and (\ref{Theorem1c}) of allowed electromagnetic fields.

In the next section the field equations in the case of stationary cyclic symmetric $(2+1)$ anti-de Sitter spacetimes  with nonlinear electrodynamics are derived, and then,  in Sec. \ref{Sec3},  restricting ourselves to the branch $c \ne 0$ and to the Lagrangian $L(F)= \sqrt{-sF}$, a family of solutions is determined.

%%%%%%%%%%%%%%%%%%%%%%%%%%%%%%%%%%%%%%%%%%%%%%%%%%%%%%%%%%%%%%%%%%
\section{The field equations for stationary and cyclic symmetric $(2+1)-$dimensional spacetime in General Relativity coupled to Nonlinear Electrodynamics}
\label{Sec2}

Starting from the following general form of the metric for stationary and cyclic symmetric (2+1) spacetime, written in the coordinates 
$\{$ $t$, $r$, $\phi$ $\}$, 
\begin{equation}\label{ScyclicMe}
ds^{2} = -N^{2}(r)dt^{2} + \frac{dr^{2}}{f^{2}(r)} + r^{2}(d\phi + \omega(r)dt)^{2},    
\end{equation}
where $N(r)$, $f(r)$ and $\omega(r)$ are functions that only depend on the radial coordinate (not  to be
confused $f(r)$ with the electromagnetic field tensor $\boldsymbol{f} = \frac{1}{2}f_{\mu\nu}dx^{\mu} \wedge dx^{\nu}$). This metric has off-diagonal components. However, one can define an orthonormal frame $\{\theta^{(0)}, \theta^{(1)}, \theta^{(2)}\}$,  which can be expanded into coordinate frame $\{t, r, \phi\}$ as, 

\begin{equation}
\theta^{(0)} = Ndt,  \quad \theta^{(1)} = \frac{dr}{f},  \quad \theta^{(2)} = r( d\phi + \omega dt),  
\end{equation}
in such a way that in this frame, the metric (\ref{ScyclicMe}) can be written in diagonal form, 

\begin{equation}\label{ScyclicMeDiag}
 ds^{2} = g_{(\alpha)(\beta)}\theta^{(\alpha)}\theta^{(\beta)} = -(\theta^{(0)})^{2} + (\theta^{(1)})^{2} + (\theta^{(2)})^{2}, 
\end{equation}
where  $g_{(a)(b)} = \eta_{(a)(b)}$  with $\eta_{(a)(b)} = {\rm diag}(-1, 1, 1)$, hence in the orthonormal frame some calculations are simplified.  
Quantities in this frame will be written with subscripts in round brackets.
Now, in order to calculate the nonvanishing components of the Einstein tensor in the orthonormal frame, we will use the Cartan's structure equations. The first Cartan's structure equation reads as,

\begin{equation}
d\theta^{(\alpha)} = - \omega^{(\alpha)}{}_{(\beta)} \wedge  \theta^{(\beta)},     
\end{equation}
where $\omega^{(\alpha)}{}_{(\beta)}$ are the connection 1-forms. Explicitly, by making use of the first Cartan's structure equation, one can find that the nonvanishing connection 1-forms are given by,
\begin{eqnarray}
&&\omega^{(0)}{}_{(1)} = \frac{fN_{,r}}{N}\theta^{(0)} - \frac{rf\omega_{,r}}{2N}\theta^{(2)}, \\ 
&&\omega^{(0)}{}_{(2)} = - \frac{rf\omega_{,r}}{2N}\theta^{(1)}, \\ 
&&\omega^{(1)}{}_{(2)} =  - \frac{rf\omega_{,r}}{2N}\theta^{(0)} - \frac{f}{r}\theta^{(2)}. 
\end{eqnarray}
which satisfy $\omega^{(i)}{}_{(j)} = -\omega^{(j)}{}_{(i)}$ and $\omega^{(0)}{}_{(j)} = \omega^{(j)}{}_{(0)}$, with $i$, $j$ $= 1 , 2$. Moreover, the comma denotes ordinary derivative with respect to the radial coordinate; $r$.
Continuing with the construction of the curvature quantities, the second Cartan's structure equation reads as, 

\begin{equation}
\frac{1}{2}R^{(\alpha)} {}_{(\beta)(\nu)(\mu)} \theta^{(\nu)} \wedge\theta^{(\mu)} = d \omega^{(\alpha)} {}_{(\beta)}  +   \omega^{(\alpha)} {}_{(\nu)} \wedge \omega^{(\nu)} {}_{(\beta)},     
\end{equation}
where $R^{(\alpha)} {}_{(\beta)(\nu)(\mu)}$ are the Riemann tensor components relative to the orthonormal frame.
By using  the second Cartan's structure equation, one can find that the nonvanishing components of the Riemann tensor relative to the orthonormal frame are given by, 

\begin{eqnarray}
&&R_{(0)(1)(0)(1)} = \frac{f}{N}\left[ (fN_{,r})_{\!,r} -\frac{ r^{2}f }{2N}(\omega_{,r})^{\!2}   \right]- \left(\frac{r\!f\omega_{\!,r}}{2N}  \right)^{2}, \\ 
&&R_{(0)(1)(1)(2)} = \frac{f}{r} \left( \frac{r^{2}f\omega_{,r}}{2N}   \right)_{\!\!\!,r} + \frac{f^{2}\omega_{,r}}{2N}, \\ 
&&R_{(0)(2)(0)(2)} =  \frac{f^{2}N_{,r}}{rN} +   \left( \frac{r f\omega_{,r}}{2N}  \right)^{2}, \\
&&R_{(1)(2)(1)(2)} =  - \frac{ff_{,r}}{r} -   \left(\frac{r f\omega_{ ,r}}{2N}  \right)^{2}.
\end{eqnarray}

The Ricci tensor $\boldsymbol{R} = R_{(\nu)(\mu)} \theta^{(\nu)} \otimes \theta^{(\mu)}$ and the curvature scalar (or the Ricci scalar) $``R"$ can be defined by the contractions, so that the Ricci tensor components relative to the orthonormal frame are calculated as $R_{(\nu)(\mu)} = \eta^{(\alpha)(\beta)} R_{(\alpha)(\nu)(\beta)(\mu)}$. Whereas the Ricci scalar, $R = \eta^{(\nu)(\mu)}R_{(\nu)(\mu)}$ is given by,

\begin{equation}
R = 2 \left [-\frac{f}{N}(fN_{,r})_{,r} +  \left( \frac{ rf\omega_{,r} }{2N}  \right)^{2} - \frac{f^{2} N_{,r}}{rN} - \frac{f f_{,r}}{r}  \right].    
\end{equation}
On the other hand, given that the Einstein tensor components relative to the orthonormal frame are %calculated 
defined as $G_{(\nu)(\mu)} = R_{(\nu)(\mu)} - \frac{R}{2}\eta_{(\nu)(\mu)}$, then explicitly, we find that the non null Einstein tensor components $G_{(\nu)(\mu)}$, are, 

\begin{eqnarray}
&&G_{(0)(0)} = -  \left( \frac{rf\omega_{,r}}{2N}  \right)^{2} -  \frac{f f_{,r}}{r}, \\ 
&&G_{(1)(1)} =  \left (\frac{rf\omega_{,r}}{2N}  \right)^{2} + \frac{f^{2} N_{,r}}{rN}, \\ 
&&G_{(2)(2)} =  \frac{f}{N}(fN_{,r})_{,r} - \frac{3}{4} \left(\frac{rf\omega_{,r}}{N}  \right)^{2},\\
&&G_{(0)(2)} =  -\frac{f}{r} \left( \frac{r^{2}f\omega_{,r}}{2N}  \right)_{,r} - \frac{f^{2}\omega_{,r}}{2N}.
\end{eqnarray}

Regarding the matter field,  in order to work with (\ref{EinsteinEqs})  with the left hand side given by $G_{(\alpha)(\beta)}$, 
the electromagnetic field should also be written in terms of the orthonormal frame.  By  comparison between $\boldsymbol{f}$ written in the frame 
$\{ dx^{\alpha} \wedge dx^{\beta} \}_{x^{\alpha},x^{\beta} = t,r,\phi}$ and the one written in  $\{ \theta^{(\alpha)} \wedge \theta^{(\beta)} \}_{(\alpha),(\beta) = 0,1,2}$,  it  is obtained that,

\begin{equation}
 \boldsymbol{f} = f_{tr} dt \wedge  dr + f_{r\phi} dr \wedge d \phi + f_{t\phi} dt \wedge  d\phi 
  = f_{(0)(1)} \theta^{(0)} \wedge \theta^{(1)} 
  + f_{(0)(2)} \theta^{(0)} \wedge  \theta^{(2)} + f_{(1)(2)} \theta^{(1)} \wedge \theta^{(2)},   
\end{equation}
from which it can be determined $f_{(a)(b)}$ in terms of $f_{ab}$; finding that the non-vanishing components of  $f_{(a)(b)}$ are, 
 
\begin{equation}
f_{(0)(1)} = -\frac{b}{rL_{F}},  \quad f_{(0)(2)} = \frac{3c}{rN}, \quad   
 f_{(1)(2)} = \frac{a-b\omega}{NL_{F}}.   
\end{equation}
On the other hand, the energy-momentum tensor components in the frame $\{ \theta^{(0)},  \theta^{(1)}, \theta^{(2)} \}$ are given by, 

\begin{equation}
4\pi E_{(\alpha)(\beta)} = \eta_{(\alpha)(\beta)} L(F) - L_{F} f_{(\alpha)(\nu)}f_{(\beta)}{}^{(\nu)}.     
\end{equation}
Thus, one finds that the non null energy-momentum tensor components in the orthonormal frame becomes, 

\begin{eqnarray}
&&4\pi E_{(0)}{}^{(0)} = - L_{F}( f_{(0)(1)} f^{(0)(1)}  +  f_{(0)(2)} f^{(0)(2)}  )  + L, \quad 
4\pi E_{(1)}{}^{(1)} = - L_{F}( f_{(1)(0)} f^{(1)(0)}  + f_{(1)(2)} f^{(1)(2)}  ) + L, \\ 
&&4\pi E_{(2)}{}^{(2)}= - L_{F}( f_{(2)(0)} f^{(2)(0)} + f_{(2)(1)} f^{(2)(1)}  )  + L, \quad 
4\pi E_{(2)}{}^{(0)} = - L_{F} f_{(2)(1)} f^{(0)(1)}, \\
&&4\pi E_{(1)}{}^{(0)} = - L_{F} f_{(1)(2)} f^{(0)(2)},\quad \quad \quad \quad \quad \quad \quad \quad \quad \quad 4\pi E_{(2)}{}^{(1)} = - L_{F} f_{(2)(0)} f^{(1)(0)}. 
\end{eqnarray}

Since $G_{(1)(2)} = 0 = g_{(1)(2)}$, then (via Einstein equations) one arrives at;  $E_{(1)(2)} = 0 \Rightarrow  f_{(2)(0)}f^{(1)(0)} = 0 \Rightarrow bc = 0.$ While, $G_{(1)(0)} = 0 = g_{(1)(0)}$ together with $bc=0$, implies  $ac=0$.  This constitutes another way to prove the vanishing conditions $ac=0=bc$, Eq. (\ref{Theorem1a}), avoiding  the exterior calculus formalism implemented in Sec. \ref{Sec1}.
%%%%%%%%%%%%%%%%%%%%%%%%%%%%%%%%%%%%%%%%%%%%%%%%%%%

\subsection{Einstein field equations in the branch $c\neq0$, $a=0=b$.}

Specifically for the branch $c\neq0$, $a=0=b$, it is obtained that $\boldsymbol{f} = f_{t\phi} dt \wedge d\phi =  f_{(0)(2)} \theta^{(0)}\wedge \theta^{(2)}$. Whereas, the field equations of general relativity (with cosmological constant) coupled to NLED in a stationary and cyclic symmetric (2+1) spacetime, written in the orthonormal frame become,    

\begin{eqnarray}
&& G_{(0)}{}^{(0)} = 8\pi E_{(0)}{}^{(0)} - \Lambda \delta_{(0)}^{(0)}  \quad  \Rightarrow  \quad \left( \frac{ rf\omega_{,r} }{2N}   \right)^{2} + 
\frac{ (f^{2})_{,r} }{2r} = 2 \left( L - 2FL_{F}  \right) - \Lambda, \label{field_C_1}\\
&&G_{(1)}{}^{(1)} = 8\pi E_{(1)}{}^{(1)} - \Lambda \delta_{(1)}^{(1)}  \quad \Rightarrow  \quad  \left( \frac{ rf\omega_{,r} }{2N}   \right)^{2} + 
\frac{ f^{2} N_{,r} }{rN} = 2L - \Lambda, \label{field_C_2}\\
&&G_{(2)}{}^{(2)} = 8\pi E_{(2)}{}^{(2)} - \Lambda \delta_{(2)}^{(2)}  \quad \Rightarrow  \quad \frac{ f (f N_{,r} )_{,r} }{N} - \frac{3}{4} \left(  
\frac{ rf\omega_{,r} }{N}   \right)^{2} = 2 \left(  L - 2FL_{F}  \right) - \Lambda, \label{field_C_3}\\
&&G_{(2)}{}^{(0)} = 8\pi E_{(2)}{}^{(0)} - \Lambda \delta_{(2)}^{(0)}  \quad \Rightarrow  \quad \frac{f}{r} \left(  \frac{ r^{2}f\omega_{,r} }{2N}   \right)_{,r} + \frac{ f^{2}\omega_{,r} }{2N} = 0, \label{field_C_4}\\
&&\nabla_{(a)}(L_{F}f^{(a)(b)}) = 0 = \nabla_{(a)}( _{\ast}\boldsymbol{f} )^{(a)}  \quad   \Rightarrow   \quad  F = -\frac{1}{2} \left( \frac{3c}{rN}  \right)^{2}. \label{field_C_5}
\end{eqnarray}
The field  Eq. (\ref{field_C_4}) implies that;

\begin{equation}\label{field_C_4_A}
r \left( \frac{ r^{2}f\omega_{,r} }{N}   \right)_{,r} + \frac{ r^{2}f \omega_{,r} }{N} = 0 \Rightarrow \frac{ f\omega_{,r} }{N} = %\textcolor{red}{-} 
\frac{J}{r^{3}},  
\end{equation}
where $J$ is an integration constant.

From the previous equations we proceed to determine $f(r), N(r), \omega(r)$ and the corresponding electromagnetic nonvanishing component, $f_{t \phi}$ for a NLED characterized by $L(F) = \sqrt{-s F}$, where $``s"$ is a positive parameter introduced  in order that $L(F)$, in the action (\ref{actionf}), has the right dimensions. %%%%%%%%%%%%%%%%%%%%%%%%%%%%%%%%%%%%%%%%%%%%%%%%%%%%%%%%%%%%

\section{New stationary and cyclic symmetric exact solution in (2+1) general relativity  coupled to Nonlinear Electrodynamics}\label{Sec3}

 We shall assume a NLED derived from $L(F) = \sqrt{-s F}$ with $s > 0$, where the invariant $F$, in the branch ($c\neq0$, $a=0=b$) is given by Eq. (\ref{field_C_5}).  Restricting ourselves to this model\footnote{ $L(F) = \sqrt{ s F}$ with $s\in \mathbb{R}$ } $(L - 2FL_{F}) = 0$ holds, and Eq. (\ref{field_C_1}) reduces to;

\begin{equation}
 \left(  \frac{ rf\omega_{,r} }{2N}   \right)^{2} + \frac{ (f^{2})_{,r} }{2r} = - \Lambda.     
\end{equation}
Now, since ${ f\omega_{,r} }/{N} = {J}/{r^{3}}$, the previous Eq. yields; 

\begin{equation}
 \frac{ (f^{2})_{,r} }{2r} + \frac{J^{2}}{4r^{4}}  = - \Lambda,     
\end{equation}
whose general solution corresponds to,

\begin{equation}\label{Sol_f}
 f^{2}(r) = - M + \frac{J^{2}}{4r^{2}}   - \Lambda r^{2},     
\end{equation}
where $M$ is an integration constant. 
Working now with Eq.(\ref{field_C_3}) which for the characteristic function $L(F) = \sqrt{-s F}$, yields, 

\begin{equation}
\frac{ f (f N_{,r} )_{,r} }{N} - \frac{3}{4} \left(  \frac{ rf\omega_{,r} }{N}   \right)^{2} = - \Lambda,    
\end{equation}
by substituting ${ f\omega_{,r} }/{N} = {J}/{r^{3}}$, and $f(r)$ from  Eq.(\ref{Sol_f}), into the previous equation, we find,

\begin{equation}
 f^{2} N_{,r,r} + \frac{(f^{2})_{,r}}{2} N_{,r}  - \frac{3}{4}\frac{ J^{2} }{ r^{4} }N  = - \Lambda N, 
\end{equation}
whose general solution is given by,

\begin{equation}\label{Sol_N}
 N(r) = q_{\alpha}   \left(  - M + \frac{J^{2}}{2r^{2}}   \right)r    +  q_{\beta} \sqrt{ - M  + \frac{J^{2}}{4r^{2}} - \Lambda r^{2} }.     
\end{equation}
where $q_{\alpha}$ and $q_{\beta}$ are two new integration constants related to the NLED. Whereas, using  Eq. (\ref{field_C_4_A}) we determine $\omega (r)$ as  the integral,

\begin{equation}\label{Sol_omega_G}
\omega(r) =
\int\frac{J}{r^{3}} \frac{N(r)}{f(r)} dr.    
\end{equation} 
Then replacing Eqs. (\ref{Sol_f}) and (\ref{Sol_N}) into  the previous equation and integrating, we find, 

\begin{equation}\label{Sol_omega}
\omega(r) = - \frac{ q_{\beta} J }{2 r^{2}} - \frac{ q_{\alpha} J }{r}\sqrt{ - M  + \frac{J^{2}}{4r^{2}} - \Lambda r^{2} }  + \omega_{\infty},  
\end{equation}
where $\omega_{\infty}$ is an integration constant. Without any loss of generality one can always set $\omega_{\infty}=0.$
Finally, working with Eq.(\ref{field_C_2}), which for the characteristic function $L(F)=\sqrt{-s F}$, becomes, 

\begin{equation}
\frac{ f^{2} N_{,r} }{rN} +  \frac{J^{2}}{r^{4}} - 2\sqrt{-s F} + \Lambda = 0   
\end{equation}
and substituting  $F(r)$, $f(r)$ and  $N(r)$ from Eqs.  (\ref{field_C_5}), (\ref{Sol_f}) and  (\ref{Sol_N}),  respectively,  we find the relation,

\begin{equation}\label{ligadura}
 \left(  M^{2} q_{\alpha} + \Lambda q_{\alpha} J^{2} - 3c\sqrt{2s}   \right) \left[ (J^{2} - 2Mr^{2}) q_{\alpha}\sqrt{ - 4Mr^{2} + J^{2} -4\Lambda r^{4} } - 4q_{\beta}\Lambda r^{4}  - 4Mq_{\beta} r^{2} + q_{\beta}J^{2}  \right] = 0. 
\end{equation}
If the parameters of the metric are chosen in order to satisfy Eq.(\ref{ligadura}) for all $r$, then $c$ is given by,  

\begin{equation}\label{Sol_c} 
c  =  \frac{1}{6}\sqrt{\frac{2}{s}}( M^{2} + J^{2}\Lambda ) q_{\alpha}, 
\end{equation}
and in this way Eq.(\ref{ligadura}) is satisfied in a trivial way.
Therefore, we have determined a five-parametric family of solutions given by, 

\begin{eqnarray}\label{NewSolution}
ds^{2} &=&  -  \left[ q_{\alpha}\left( - M + \frac{J^{2}}{2r^{2}}\right)r    +  q_{\beta} \sqrt{ - M  + \frac{J^{2}}{4r^{2}} - \Lambda r^{2}}  \right]^{2} dt^{2} + \frac{dr^{2}}{ - M + \frac{J^{2}}{4r^{2}}  - \Lambda r^{2} } \nonumber \\
&& + r^{2}\left[d\phi - J\left( \frac{ q_{\beta} }{2 r^{2}} + \frac{ q_{\alpha} }{r}\sqrt{- M  + \frac{J^{2}}{4r^{2}} - \Lambda r^{2}} \right) dt \right]^{2}. 
\end{eqnarray}
%%%
with parameters  $M$, $J$, $q_{\alpha}$, $q_{\beta}$ and $\Lambda$. For this metric the Ricci scalar is given by,

\begin{equation}\label{RicciScalar}
R(r) = 6\Lambda - \frac{ 2q_{\alpha}(M^{2} + J^{2}\Lambda)}{rN(r)}.    
\end{equation}
On another hand, in order to verify the consistency between the  system of field equations with the metric  (\ref{NewSolution}), we calculate the trace of Einstein Eqs.(\ref{EinsteinEqs}),  this is,

\begin{equation}
g^{(\alpha)(\beta)}G_{(\alpha)(\beta)} = 8\pi g^{(\alpha)(\beta)} E_{(\alpha)(\beta)} - \Lambda g^{(\alpha)(\beta)}g_{(\alpha)(\beta)}\quad \Rightarrow\quad 3L - 4FL_{F} + \frac{R}{4} - \frac{3}{2} \Lambda = 0.
\end{equation}
Which for the model $L(F) = \sqrt{-s F} $, yields, 

\begin{equation}\label{Scalartrace}
\sqrt{-s F(r)} + \frac{R(r)}{4} - \frac{3}{2} \Lambda = 0.     
\end{equation}
Now, by using Eq. (\ref{field_C_5}) and solving for $R$, we find that; 

\begin{equation}\label{RicciScalarc}
R(r) =  6\Lambda - 4\sqrt{-s F(r)} = 6\Lambda -   \frac{3c\sqrt{8s}}{rN(r)}.        
\end{equation}
Finally, substituting $c$, Eq. (\ref{Sol_c}), into the previous equation, it is obtained,

\begin{equation}
R(r) =  6\Lambda - 4\sqrt{-s F(r)} = 6\Lambda -  \frac{2q_{\alpha}(M^{2} + J^{2}\Lambda )}{rN(r)},  
\end{equation}
that agrees with Eq. (\ref{RicciScalar}), obtained directly from the metric.

Briefly we analyze the zeroes of $N(r)$ and $f(r)$, the metric functions in  Eq. (\ref{NewSolution}).
$N^{2}(r)$ and $f^{2}(r)$ do not have the same roots. In general, the roots  of $N(r)$ (i.e., the solutions of the equation $N^{2}(x)=0$) and the roots of  $f^{2}(r)$ (the solutions of the equation $f^{2}(x)=0$), respectively, are given by,

\begin{eqnarray}
&& x_{_{1,N}} = \sqrt{ \frac{M (J^{2}q_{\alpha}^{2}-  q_{\beta}^{2} )+  \sqrt{ q_{\beta}^{2}( M^{2}+  \Lambda J^{2} ) ( q_{\beta}^{2}- J^{2}q_{\alpha}^{2}) } }{ 2(\Lambda q_{\beta}^{2}+\! M^{2}q_{\alpha}^{2}) } }, \label{x1N}\\
&&x_{_{2,N}} = \sqrt{\frac{M (J^{2}q_{\alpha}^{2}-\! q_{\beta}^{2}\!) - \sqrt{q_{\beta}^{2}(M^{2}+ \Lambda J^{2} ) ( q_{\beta}^{2}- J^{2}q_{\alpha}^{2}\!)  } }{ 2(\Lambda q_{\beta}^{2}+\! M^{2}q_{\alpha}^{2}) } }, \label{x2N}\\
&&x_{_{1,f}} = \sqrt{ -\frac{1}{2\Lambda}  \left( M+ \sqrt{M^{2}+ \Lambda J^{2}}   \right) },\quad \textup{  with  } \Lambda \neq0 \label{x1f}\\ 
&&x_{_{2,f}} = \sqrt{ -\frac{1}{2\Lambda} \left( M- \sqrt{M^{2}+ \Lambda J^{2}}    \right) },\quad  \textup{  with  } \Lambda \neq0.\label{x2f}
\end{eqnarray}
where the subscript indicates if the root comes from  $N=0$ or $f=0$. In a first thought we could think of these zeroes as being horizons. However, the
reason why these roots are not event horizons is that at $r=x_{_{1, N}}$ and $r=x_{_{2, N}}$, according to Eq. 
(\ref{RicciScalar}), the Ricci scalar
diverges in these regions, and then they correspond to curvature singularities. This is the main reason why, in general, the metric (\ref{NewSolution}) cannot be interpreted as a (2+1)-dimensional black hole. However, under certain restrictions $R$ is finite, as we shall show in what follows.

%%%%%%%%%%%%%%%%%%%%%%%%%%%%%%%%%%%%%%%%%%%%%%%%%%%%%%%%%

\subsection{Uncharged limit ($q_{\alpha}=0$); the BTZ solution }\label{btzBH}

By switching off the electromagnetic parameter $q_{\alpha}$, holding $q_{\beta}$ different from zero,
assuming  $M$ be positive and $\Lambda$ be negative,  in the derived metric  Eq. (\ref{NewSolution}), we recover the BTZ black hole,

\begin{equation}\label{uncharged1}
ds^{2} =  -   \left(- M  + \frac{J^{2}}{4r^{2}} - \Lambda r^{2}   \right)  (q_{\beta}dt)^{2} + \frac{dr^{2}}{- M + \frac{J^{2}}{4r^{2}}   - 
\Lambda r^{2} } + r^{2} \left( d\phi -  \frac{ J }{2 r^{2}}  (q_{\beta}dt )\right)^{2}.
\end{equation}
Now, renaming $\tilde{t} \rightarrow q_{\beta} t$, the metric (\ref{uncharged1}) yields,

\begin{equation}
ds^{2} =  -   \left( - M  + \frac{J^{2}}{4r^{2}} - \Lambda r^{2}   \right)  d\tilde{t}^{2} + \frac{dr^{2}}{ - M + \frac{J^{2}}{4r^{2}}   - \Lambda r^{2} } + r^{2} \left( d\phi - \frac{ J }{2 r^{2}} d\tilde{t} \right)^{2}
\end{equation}
that can be identified as the BTZ black hole \cite{BTZ1992} with $M$ being the mass, $J$ the angular momentum and $\Lambda$ the anti de Sitter parameter, sometimes parametrized as $\Lambda=-1/l^2$.

%%%%%%%%%%%%%%%%%%%%%%%%%%%%%%%%%%%%%%%%%%%%%%%%%%%%%

\subsection{Black hole with non-null electromagnetic charges $q_{\alpha}$ and $q_{\beta}$, but with vanishing electromagnetic field; $f_{(a)(b)} =0.$}\label{newextremeBH} 

We can restrict the parameters of the five-parametric family, Eq. (\ref{NewSolution}), in order that the solution does possess event horizons, $r_{h}$. Defining the event horizons by $N(r_{h}) = 0 = f(r_{h})$, this condition leads to,

\begin{equation}
- M + \frac{J^{2}}{2r^{2}_{h}}  = 0 = - M  + \frac{J^{2}}{4r^{2}_{h}} - \Lambda r^{2}_{h}.
\end{equation}
The first equation is satisfied with  $r^{2}_{h} = {J^{2}}/{ 2 M }$, with $M>0$ in order that $r_{h}$ be real. On the other side, substituting $r^{2}_{h}$ into $M - \frac{J^{2}}{4r^{2}_{h}} + \Lambda r^{2}_{h} = 0$, we get the relation; $\Lambda = - { M^{2} }/{ J^{2} }$, this value of $\Lambda$ we shall call $\Lambda_0$. Checking the self-consistency of our equations, substituting $\Lambda = - { M^{2} }/{ J^{2} }$ into the Eqs. (\ref{x1N}), (\ref{x2N}), (\ref{x1f}) and (\ref{x2f}), the following roots are obtained,

\begin{equation}
x^{2}_{_{1,N}} = x^{2}_{_{2,N}} = x^{2}_{_{1,f}} = x^{2}_{_{2,f}} = \frac{J^{2}}{ 2 M }, 
\end{equation}
i.e., the equations $N^{2}(x_{_{N}})=0$ and $f^{2}(x_{_{f}})=0$ have only one positive root, and this positive root is 
the same for both functions. 
%%%
Therefore, in  this case ($\Lambda_0 = - { M^{2} }/{ J^{2} }$), the metric  (\ref{NewSolution}) yields,

\begin{equation}\label{newBH}
ds^{2}  =  - \left(Jq_{\alpha} + q_{\beta}\right)^{2}\left( - M + \frac{J^{2}}{2r^{2}}\right)^{2}\frac{r^{2}}{J^{2}} dt^{2} + \frac{dr^{2}}{ \frac{ r^{2} }{ J^{2} } \left( -  M + \frac{ J^{2} }{ 2r^{2} } \right)^{2}   }  + r^{2}\left\{d\phi - \left[ \left( Jq_{\alpha} +  q_{\beta}  \right)\frac{ J }{ 2r^{2} } -  q_{\alpha} M \right]  dt \right\}^{2}.    
\end{equation}
Besides, substituting $\Lambda = \Lambda_0 = - { M^{2} }/{ J^{2} }$ into Eq. (\ref{Sol_c}), for $c$, it is found,  

\begin{equation}\label{Sol_cTilde} 
c  =  \frac{1}{6}\sqrt{\frac{2}{s}}( M^{2} + J^{2}\Lambda_{0} ) q_{\alpha} = 0, 
\end{equation}
that from Eq. (\ref{field_C_5}) implies $F=0$, and $ \boldsymbol{f} = f_{(0)(2)} d\theta^{(0)}\wedge d\theta^{(2)} =0$.
Then, evaluating the energy-momentum tensor for this configuration  we obtained, $E_{ab}$ = 0. 

On the other hand, by using (\ref{RicciScalar}) or (\ref{Scalartrace}), we can check that for the metric (\ref{newBH}) the Ricci scalar curvature is constant  for all $r$, 

\begin{equation}\label{RicciSnewVS}
R(r) = -\frac{ 6 M^{2} }{ J^{2} } = 6\Lambda_0. 
\end{equation}
Therefore the Ricci scalar is regular everywhere. In particular, in contrast to the case $c\neq0$, the curvature scalar (\ref{RicciSnewVS}) is regular in the whole range of $r$.
 Moreover, through straightforward calculation using (\ref{newBH}), one can find that the line element  has the following scalar invariants\footnote{  For (2+1)-GR with cosmological constant and null energy momentum tensor; $G_{ab} = R_{ab} - \frac{1}{2} Rg_{ab} = -\Lambda g_{ab}$ implies  $R=6\Lambda$, thus $R_{ab} =  2\Lambda g_{ab}$ then $R_{ab}R^{ab} = 12\Lambda^{2}$. Besides, in (2+1)-dimensional spacetime, the full curvature tensor is completely determined by the Ricci tensor,
$R_{abcd} = g_{ac}R_{bd} + g_{bd} R_{ac} - g_{bc} R_{ad} - g_{ad} R_{bc} - \frac{1}{2}(g_{ac}g_{bd} - g_{ad}g_{bc})R,$
this implies $R_{abcd}R^{abcd} = 12\Lambda^{2}$.},

\begin{eqnarray}
&&R_{(\alpha)(\beta)}R^{(\alpha)(\beta)} = \frac{ 12 M^{4} }{ J^{4} }, \\
&&R_{(\alpha)(\beta)(\nu)(\mu)}R^{(\alpha)(\beta)(\nu)(\mu)} = \frac{ 12M^{4} }{ J^{4} }.
\end{eqnarray}
From these expressions it  is  clear that $r = r_{h}$  is not a  curvature singularity. Furthermore, for the solution (\ref{newBH}) the surface gravity turns out to be zero.
It is easy to calculate the surface gravity, $\kappa$ from the definition \cite{Brown1994},

\begin{equation}
\kappa= - \frac{1}{2} \frac{\partial_{r} g_{tt}}{\sqrt{-g_{tt} g_{rr}}},
\end{equation}
that for the metric  (\ref{newBH})  becomes,

\begin{equation}
\kappa= \frac{q_{\beta} M^2 (-J^2+2Mr^2)(2 q_{\alpha}J + q_{\beta})}{2 J^2 \sqrt{M q_{\beta}[-J^2 (Jq_{\alpha} + q_{\beta}) + Mr^2  (2Jq_{\alpha}+q_{\beta}) ]}}, 
\end{equation}
expression that vanishes when evaluated at $r^{2}_{h} = {J^{2}}/{ 2 M }$.
Therefore the black hole is an extreme one with event horizon at $r_{h} = \frac{|J|}{ \sqrt{2M} }.$

We remark that  the solution (\ref{newBH}) is not a hairy black hole one: 
as noted following Eq. (\ref{Sol_cTilde}), the metric (\ref{newBH}) is a vacuum solution of three dimensional general relativity (with a negative cosmological constant). 
Besides,   the parameters $q_{\alpha}$ and $q_{\beta}$ cannot be considered as NLED-BH hair,  because the metric (\ref{newBH})  reduces to the extreme BTZ solution by accomplishing the coordinate transformations,
\begin{equation}
\tilde{\phi} = \phi + M q_{\alpha}t, \quad \tilde{t} = \left( Jq_{\alpha} +  q_{\beta}  \right) t, \quad \tilde{r} = r. 
\end{equation} 
where  the tilde,  $\tilde{}$, denotes the corresponding  BTZ coordinates.

%%%%%%%%%%%%%%%%%%%%%%%%%%%%%%%%%%%%%%%%%%%%%%%%% 
\subsection{Five-parametric Wormhole supported by NLED plus $\Lambda$ }\label{WHfNLED} 

The construction of wormhole structures supported only by NLED, have been analyzed in different works, see for instance Refs. \cite{Bronnikov2017,LoboArella2006,tesis} in which non-stationary axisymmetric traversable wormholes, supported only by nonlinear electrodynamics, were built.  Alternatively,  in Ref. \cite{Hendi}  the cut-and-paste method was used in order to construct a wormhole structure. We shall prove in the following that the previously derived solution (\ref{NewSolution}), with no need to resort to the cut-and-paste method, admits a stationary axisymmetric traversable WH interpretation. 

In the first part of this subsection we enumerate the wormhole properties, and then we shall prove that the metric (\ref{NewSolution})
 satisfy these conditions provided the parameters are restricted to satisfy some inequalities.

The standard form of the line element corresponding to a stationary and circularly symmetric (2+1) dimensional  wormhole  is given by,

\begin{equation}\label{whAnsatz}
ds^{2} = -e^{2\Phi(r)} dt^{2} + \frac{dr^{2}}{ 1 - \frac{b(r)}{r}}  + r^{2} [ d\phi + \omega(r)dt ]^{2}, 
\end{equation}
where, according to \cite{ThorneMorris},  $\Phi(r)$ and $b(r)$ are functions of the radial coordinate $``r"$. $\Phi(r)$ is called the red-shift function, for it is related to the gravitational redshift; whereas $b(r)$ is denoted as the shape function.  The radial coordinate has a range that increases from a minimum value at $r_{0}$, corresponding to the wormhole throat, $b(r_{0}) = r_{0}$, to $r \rightarrow \infty$. Therefore, the radial coordinate $r$ has a special geometric significance, where $2\pi r$  is the circumference of a circle centered on the wormhole throat.    
On the other hand, for the wormhole to be traversable, one must demand the absence of event horizons, which are identified as the curves  
where $e^{2\Phi(r)} \rightarrow 0$.  If $\Phi(r)$ is a continuous, nonvanishing and finite function in the whole range of $r$,  $r\in[r_{0},\infty)$, there will not occur horizons. Moreover, a fundamental property of traversable wormholes is the fullfilment of the flaring out condition, which is deduced from the mathematics of embedding, and is given by $(b - rb_{,r})/b^{2} > 0$. Note that at the throat,  $b(r_{0}) = r_{0},$  the flaring out condition reduces to $b_{,r}(r_{0})<1$. The condition $(1 - b/r) \geq 0$ is also imposed for all values of $r$.

Furthermore considering a null vector in the orthonormal frame, $\boldsymbol{n} = (1,1,0) = n^{(\alpha)}e_{(\alpha)}$,  where $\{ n^{(\alpha)} \}_{\alpha = 0}^{2}$ are the components of the null vector in the orthonormal basis vectors given by $\{ e_{(\alpha)} \}_{\alpha = 0}^{2}$ 
%with $e_{(\alpha)}$ 
such that $\theta^{(\alpha)}e_{(\beta)} = \delta^{(\alpha)}_{(\beta)}$; (we can see that  $ \boldsymbol{n}$ is a null vector since $ ds^{2}( \boldsymbol{n}, \boldsymbol{n}) = g_{(\alpha)(\beta)}n^{(\alpha)}n^{(\beta)} = 0$), 
then contracting the Einstein tensor with $ \boldsymbol{n}$ we find,

\begin{equation}
G_{(\alpha)(\beta)} n^{(\alpha)} n^{(\beta)} =
\frac{1}{r} \left( 1 - \frac{b(r)}{r}  \right) \Phi_{,r}(r) - \frac{1}{2r^{3}} \left[  b(r) - rb_{,r}(r)  \right], 
\end{equation}
equation that when evaluated at the throat,  $r=r_{0}$, and using the fact that  $\Phi(r)$ is continuous and finite everywhere, then $G_{(\alpha)(\beta)} n^{(\alpha)} n^{(\beta)}|_{r=r_{0}}$ reduces to, 

\begin{eqnarray}
 G_{(\alpha)(\beta)}n^{(\alpha)}n^{(\beta)}  \Big|_{r=r_{0}} =  -\frac{1}{2r^{3}_{0}} \left[ b( r_{0}) - r_{0}b_{,r}(r_{0})  \right] =\frac{1}{2r^{2}_{0}} \left[ b_{,r}
(r_{0})- 1  \right]< 0, \quad  {\rm  since } \quad  b_{,r}(r_{0}) < 1.
\end{eqnarray}
Now, by using the Einstein field equations and since $ \boldsymbol{n}$ is a null vector, one can conclude that
the Null Energy Condition (NEC),  which establishes that $E_{(\alpha)(\beta)}n^{(\alpha)}n^{(\beta)}\geq0$  for any null vector $n^{(\alpha)}$, is violated,  since at the wormhole throat $r_{0}$, for $ \boldsymbol{n} = (1,1,0)$, it  is obtained that, 

\begin{equation}
E_{(\alpha)(\beta)}n^{(\alpha)}n^{(\beta)}\!\Big|_{r=r_{0}} < 0.     
\end{equation}
%%%%%
Given that  the flaring out condition implies the  violation of the Null Energy Condition, then the violation of the NEC is considered  fundamental for a traversable wormhole.

Below, starting from the line element (\ref{NewSolution}) we shall proceed to the construction of a stationary and cyclic symmetric (2+1)-dimensional traversable wormhole supported by NLED plus $\Lambda$, with characteristic function $L(F) = \sqrt{-sF}$.  To begin with the construction,  by comparison between the line elements (\ref{whAnsatz}) and  (\ref{NewSolution}),  we obtain for the red-shift function, 

\begin{equation}\label{RedShift}
e^{2\Phi(r)} =  \left[ q_{\alpha} \left( - M+ \frac{J^{2}}{2r^{2}}  \right)r   +  q_{\beta} \sqrt{-M +  \frac{J^{2}}{4r^{2}} - \Lambda r^{2} }  \right]^{2},    
\end{equation}
while for the shape function one finds, 

\begin{equation}\label{Shape}
1 - \frac{b(r)}{r} = - M + \frac{J^{2}}{4r^{2}}  - \Lambda r^{2}.     
\end{equation}
It is possible to restrict the parameters; $M$, $J$, $q_{\alpha}$, $q_{\beta}$ and $\Lambda$
in such a way that the line element (\ref{NewSolution}) represents a stationary and cyclic symmetric  %axisymmetric
(2+1)-dimensional wormhole.
 
Note that if we substitute $\Lambda = \Lambda_0 = - {M^{2}}/{J^{2}}$ into the shift and shape functions $e^{2\Phi(r)}$ and $1 - \frac{b(r)}{r}$, respectively,   both vanish when are evaluated at $r= {|J|}/{ \sqrt{2M} }$.  For this reason it is  required that $M^{2} +J^{2}\Lambda \neq 0$ in order to have a meaningful wormhole solution. 

On the other hand, setting $\Lambda < 0$, the function $1 - \frac{b(r)}{r}$ have real positive roots, different from $x_{_{1,f}} = x_{_{2,f}} = {|J|}/{ \sqrt{2M}}$, according to (\ref{x1f}) and (\ref{x2f}).  Therefore  it is required that $M^{2} +J^{2}\Lambda > 0$ in order that $x_{_{1,f}}$ and  $x_{_{2,f}}$ in (\ref{x1f}) and (\ref{x2f}) become positive real numbers.  In this case then,  $x_{_{1,f}} >  x_{_{2,f}}$  and  it is $x_{_{1,f}}$ that plays the role of the wormhole throat $r_{0}$. This is,

\begin{equation}\label{throat}
r_{0} = \sqrt{ -\frac{1}{2\Lambda}  \left( M+ \sqrt{M^{2}+ \Lambda J^{2}}   \right) },  \quad {\rm   with }  \quad  M>0,  \quad  \Lambda < 0  \quad 
 {\rm   and   }  \quad M^{2}+ \Lambda J^{2} > 0.    
\end{equation}

Regarding now the condition that the red-shift function be finite in the range $r\in [r_{0},\infty)$, considering (\ref{x1N}) and (\ref{x2N}), one finds that a simple  way to guarantee that  $e^{2\Phi(r)} = N^{2}(r)$ does not have real roots is by imposing that $q_{\beta}^{2} - J^{2}q_{\alpha}^{2} <0$. Thus $x_{_{1,N}}$ and  $x_{_{2,N}}$, Eqs. (\ref{x1N}) and (\ref{x2N}), become complex numbers. 
Therefore, by imposing that the parameters $M$, $J$, $q_{\alpha}$, $q_{\beta}$ and $\Lambda$ satisfy,

\begin{equation}\label{WHcondition}
M>0, \quad \Lambda<0,  \quad M^{2}+ \Lambda J^{2} > 0  \quad {\rm and }  \quad  q_{\beta}^{2}- J^{2}q_{\alpha}^{2} < 0,    
\end{equation}
it is achieved that  $\Phi(r),$ defined in Eq. (\ref{RedShift}),  is  a continuous, nonvanishing and finite function in the whole range of  $r$,  $r \geq r_{0}$. Whereas the function $b(r)$,  Eq.  (\ref{Shape}),  when evaluated at the wormhole throat $r_{0}$ satisfies that $b(r_{0}) = r_{0}$, together with $1-\frac{b(r)}{r} \geq 0$  for all $r \in [r_{0},\infty)$. 

In order that the wormhole be traversable we must ensure that the flaring out condition, $[b(r) - rb_{,r}(r)] > 0,$  is fulfilled. Then for $b(r)$, Eq.(\ref{Shape}), we find that,

\begin{equation}\label{flaringcondition}
b(r) - rb_{,r}(r) = -\frac{J^{2}}{2r} - 2\Lambda r^{3}.     
\end{equation}
It can  be seen that $[b(r) - rb_{,r}(r)]$ is an increasing function on $r$ since its derivative,

\begin{equation}\label{derivadaFlaring}
[b(r) - rb_{,r}(r)]_{,r} = \frac{J^{2}}{2r^{2}} - 6\Lambda r^{2} >0,  \quad {\rm  since } \quad \Lambda <0.     
\end{equation}
Then evaluating $[b(r) - rb_{,r}(r)]$  at the minimum of the radial coordinate $r=r_{0}$, we find, 

\begin{equation}\label{flaringr0}
b(r_{0}) - r_{0}b_{,r}(r_{0})= \frac{ \sqrt{2}[ M^{2}+ \Lambda J^{2}+ M\sqrt{M^{2}+ \Lambda J^{2}}]  }{  \sqrt{-\Lambda  \left(  M+ \sqrt{M^{2}+ \Lambda J^{2}}   \right) } },     
\end{equation}
which is positive since $M > 0$, $\Lambda<0$, and  $M^{2}+  \Lambda J^{2}  > 0$. Then, from (\ref{flaringr0}) together with (\ref{derivadaFlaring}) one can conclude that  $[b(r) - rb_{,r}(r)] > 0$ for all $r \in [r_{0},\infty)$.
Thus, we have proved that the metric (\ref{NewSolution}), with restricted parameters according to  (\ref{WHcondition}), satisfies the flaring out condition.
 
The final step to complete the construction of the traversable wormhole, is to check the violation of the Null Energy Condition.  
To this end, let us consider the null vector $\boldsymbol{n} = (1,1,0)$ and let us calculate $E_{(\alpha)(\beta)}n^{(\alpha)}n^{(\beta)}$. Then, taking into account that for $L=\sqrt{-sF}$ $\Rightarrow$ $L - 2FL_{F} = 0$, and by using the right hand side of Eqs. (\ref{field_C_1}) and (\ref{field_C_2}), we arrive to, 

\begin{equation}\label{NECtoL}
8\pi E_{(\alpha)(\beta)}n^{(\alpha)}n^{(\beta)} = 2L = 2\sqrt{-sF}. 
\end{equation}
Now, based on the relation (\ref{RicciScalarc}) and the identification $N^{2}(r)= e^{2\Phi(r)}$, we find that, 

\begin{equation}\label{NECV}
2\sqrt{-s F} = \frac{3c\sqrt{4s}}{r e^{\Phi(r)} } \quad  \Rightarrow \quad  E_{(\alpha)(\beta)}n^{(\alpha)}n^{(\beta)} = \frac{3c\sqrt{s}}{4\pi r e^{\Phi(r)}},  
\end{equation}
from the previous expression it can be seen that for the case in which the parameter $c$ is positive the NEC is violated if $e^{\Phi(r)}$  becomes negative. Whereas for the case in which $c$ is negative, then, the NEC is violated if $e^{\Phi(r)}$ becomes positive.  Let us analyze both possibilities:

\begin{itemize}

\item{{\it Case} $c>0$. }

From Eq.(\ref{Sol_c}), and given that $M^{2} + J^{2}\Lambda > 0$, then  $c>0$ implies $q_{\alpha} > 0$; and  $e^{\Phi(r_{0})}$ becomes,

\begin{equation}\label{NECviol}
e^{\Phi(r_{0})} =  q_{\alpha} \left( - M+\! \frac{J^{2}}{2r^{2}_{0}}  \right) r_{0},
\end{equation}
substituting $r_{0}$ into Eq. (\ref{throat}), yields,

\begin{equation}\label{NECviola}
e^{\Phi(r_{0})} =  -  \frac{q_{\alpha} ( M^{2} + \Lambda J^{2} + M \sqrt{ M^{2} + \Lambda J^{2}  } ) }{ \sqrt{ -2 \Lambda (M + \sqrt{ M^{2} + \Lambda J^2})}}
\end{equation}
This means  $e^{\Phi(r_{0})}<0$,  since $M > 0$, $q_{\alpha}>0$, $\Lambda<0$, and $M^{2}+  \Lambda J^{2}  > 0$. Therefore,  for these values of the parameters one finds that the NEC is violated. Thus in the case  $c>0$, for the parameters fulfilling 

\begin{equation}\label{WormHoleconditionA}
M > 0,  \quad q_{\alpha}>0,  \quad \Lambda<0,  \quad M^{2}+  \Lambda J^{2}  > 0,  \quad  {\rm and } \quad  q_{\beta}^{2}-  J^{2}q_{\alpha}^{2} < 0,  
\end{equation}
the five parametric  family  of stationary and circularly symmetric (2+1) given by the metric  (\ref{NewSolution}) represents a traversable wormhole. 

\item{{\it Case} $c<0$.} 

From Eq.(\ref{Sol_c}), and given that $M^{2} + J^{2}\Lambda > 0$, then $c<0$ implies $q_{\alpha} < 0$. Now from Eq. (\ref{NECviola}) it is found that $e^{\Phi(r_{0})}$ is positive,  since $M > 0$, $q_{\alpha}<0$, $\Lambda<0$, and $M^{2}+  \Lambda J^{2}  > 0$. Therefore according to (\ref{NECV}) evaluated at $r_{0}$, for these values of the parameters one finds that the NEC is violated; $E_{(\alpha)(\beta)}n^{(\alpha)}n^{(\beta)}|_{r_{0}}<0$. Thus  in the case  $c<0$, for  parameters fulfilling,

\begin{equation}\label{WormHoleconditionB}
M > 0,  \quad q_{\alpha}<0,  \quad \Lambda<0,  \quad M^{2}+ \Lambda J^{2} > 0, \quad {\rm and } \quad q_{\beta}^{2}- J^{2}q_{\alpha}^{2} < 0.     
\end{equation}
the solution (\ref{NewSolution}) also corresponds to a five parametric family of stationary and circularly symmetric (2+1) traversable wormhole metric.

\end{itemize}

Combining the results (\ref{WormHoleconditionA}) and (\ref{WormHoleconditionB}) it is concluded that the solution (\ref{NewSolution}) is a five-parametric family of stationary and circularly symmetric (2+1) traversable wormhole provided the parameters  are restricted to satisfy,

\begin{equation}\label{WormHolecondition}
M > 0,  \quad \Lambda<0,  \quad M^{2}+  \Lambda J^{2} > 0,  \quad {\rm and } \quad  q_{\beta}^{2}-  J^{2} q_{\alpha}^{2} < 0.     
\end{equation}

In particular,  the condition $q_{\beta}^{2}- J^{2}q_{\alpha}^{2} < 0$ implies  that $q_{\alpha}\neq0$. However, we showed  in \ref{btzBH} that if the parameters fulfill  $\{ M > 0,  \quad q_{\alpha} = 0,  \quad \Lambda<0,  \quad M^{2} + \Lambda J^{2} \geq 0 \}$ then (\ref{NewSolution}) corresponds to the  BTZ black hole. Whereas in \ref{newextremeBH} we showed  that if the parameters satisfy $\{ M > 0,\quad\Lambda = -M^{2}/J^{2} \}$ for any $q_{\alpha}$ and $q_{\beta}$, then  (\ref{NewSolution})  corresponds to %a four-parametric family of stationary and circularly symmetric (2+1) 
the extreme BTZ black hole.

The previously derived stationary wormhole solution is a counterexample to the proof given in \cite{Lobo2006} about the non-existence of  (2+1)-dimensional stationary, axisymmetric traversable wormholes supported by nonlinear electrodynamics. 
In fact the conclusion of non-existence in \cite{Lobo2006} is derived from the assumed form of the electromagnetic field tensor, that is not the most general one can consider, as the form we presented in the Theorem 1, Eqs. (\ref{Theorem1}), subject to the condition (\ref{Theorem1a}).
Concretely, in the notation of Ref. \cite{Lobo2006}, their proof focuses on the stress-energy tensor components $T_{tr}$ and $T_{r\phi}$ (see Eqs. (56) and (57) in Ref. \cite{Lobo2006}). Subsequently, based on the vanishing of these stress-energy tensor components (via Einstein equations) it is concluded that the metric should satisfy $(g^{t\phi})^{2}= g^{tt}g^{\phi\phi}$ (Eq. (58) in Ref. \cite{Lobo2006}). Finally, this last condition leads to $N=0$, implying the presence of an event horizon, being this the proof of non-existence of (2+1)-dimensional stationary and axially symmetric traversable wormholes coupled to nonlinear electrodynamics. In contrast, if it is  considered the general form of the electromagnetic field tensor given by Eqs. (\ref{Theorem1}), for (2+1)-dimensional stationary and axisymmetric spacetimes, jointly with the condition Eq. (\ref{Theorem1a}), as we have assumed in the present paper, then it is found that the equations; $T_{tr} = 0$ and $T_{r\phi} = 0$ are satisfied in a trivial way, needlessly of imposing any new constraint to the metric. Therefore, solutions of the type (2+1)-dimensional stationary axisymmetric traversable wormhole supported by NLED are not ruled out. 

%%%%%%%%%%%%%%%%%%%%%%%%%%%%%%%%%%%%%%%%%%%%%%%%%%%%%%%%%%%%%%%%%%%%
\subsection{Transition Black Hole--Wormhole in Variable ``Cosmological Constant" models}\label{transitionsBHWH}

In this subsection we analyze the effect of varying the value of $\Lambda_0$ in the black hole studied in  IV B. \\
Many modern theories in order to solve cosmological puzzles propose that the fundamental constants could be dependent on time, position or the local density of matter (see, e.g. \cite{VFC,VFC1}). In particular, there are some models in which the only fundamental constant that is variable\footnote{Specifically, it is assumed that $\Lambda$ is time-dependent} is the cosmological constant $\Lambda$. These models are called Variable ``Cosmological Constant" models (VCC), see for instance  \cite{Capo1997,Overduin1998,Liu2001,Yin2008,Azri2017}. 
  
In the context of VCC, let us consider a scenario characterized by two consecutive cosmological epochs; ``$epoch$ $one$" and ``$epoch$ $two$", in which $\Lambda$ takes a constant value but different for each cosmological epoch. i.e., 
the cosmological constant behaves as a step-function  with its value depending on the cosmological epoch. This is, in the epoch one; $\Lambda = \Lambda_{0}$. Whereas in the epoch two; $\Lambda = \Lambda_{0} + \delta^{2}$. 
Schematically we will represent the transition of the cosmological constant as;
$\Lambda: \Lambda_{0} \rightarrow \Lambda_{0} + \delta^{2}$ with $\delta$ a real number such that $\delta^{2} < |\Lambda_{0}|$. \\
%%%
Let us assume the epoch one with $\Lambda = \Lambda_{0} = - {M^{2}}/{J^{2}}$,  and a spacetime given by (\ref{NewSolution}), with $M>0$ and $q_{\beta}^{2} -  J^{2}q_{\alpha}^{2}< 0$, this is, 

\begin{eqnarray}\label{bHLo}
ds^{2}&=&-  \left[q_{\alpha} \left( - M+ \frac{J^{2}}{2r^{2}}  \right) r   +  q_{\beta} \sqrt{ -M +  \frac{J^{2}}{4r^{2}}- \Lambda_{0} r^{2}}  \right]^{2} dt^{2} + \frac{dr^{2}}{- M+ \frac{J^{2}}{4r^{2}} - \Lambda_{0} r^{2} } \nonumber \\ 
&& + r^{2}\left[ d\phi- J \left( \frac{ q_{\beta} }{2 r^{2}}+ \frac{ q_{\alpha} }{r} \sqrt{- M +\! \frac{J^{2}}{4r^{2}}- \Lambda_{0} r^{2} }\right) dt\right]^{2}\!,  \!\quad {\rm with }\! \quad\! \Lambda_{0} = - \frac{M^{2}}{J^{2}},\quad \! M > 0,  \quad \! q_{\beta}^{2}- J^{2}q_{\alpha}^{2}< 0.
\end{eqnarray} 

Then, according to the discussion in  \ref{newextremeBH},  the solution (\ref{bHLo}) corresponds to %a four-parametric  stationary and circularly symmetric (2+1) 
the extreme BTZ black hole metric, with event horizon at $r_{h} = |J|/\sqrt{2M}$. 

Now suppose that the cosmological epoch change occurs. In  epoch two, $\Lambda = \Lambda_{0} + \delta^{2}$ and the metric (\ref{bHLo}) transmute to the following one,

\begin{eqnarray}\label{WHLo}
ds^{2}&=& -  \left[ q_{\alpha}\left( - M+ \frac{J^{2}}{2r^{2}}  \right)r    +  q_{\beta} \sqrt{-M + \frac{J^{2}}{4r^{2}} -  (\Lambda_{0} + \delta^{2})r^{2}}  \right]^{2} dt^{2} + \frac{dr^{2}}{- M+ \frac{J^{2}}{4r^{2}} - (\Lambda_{0}+ \delta^{2})r^{2} } + \nonumber \\ 
&& r^{2} \left[d\phi- J \left( \frac{ q_{\beta} }{2 r^{2}}+ \frac{ q_{\alpha} }{r} \sqrt{- M + \frac{J^{2}}{4r^{2}} -  (\Lambda_{0} + \delta^{2})r^{2} }\right) dt \right]^{2}\!\!,\quad \!\!\!{\rm with } \!\!\! \quad \!  M > 0, \! \quad \!\!\! q_{\beta}^{2} -  J^{2}q_{\alpha}^{2}< 0, \!\! \quad \delta^{2}< |\Lambda_{0}|.
\end{eqnarray}
Hence, $\Lambda = - {M^{2}}/{J^{2}} + \delta^{2} < 0$ since $\frac{M^{2}}{J^{2}}>\delta^{2}$. Whereas, $M^{2}+ \Lambda J^{2} =M^{2}+ (-\frac{M^{2}}{ J^{2}} + \delta^{2}) J^{2} = \delta^{2} J^{2} > 0$.

Thus, the parameters of (\ref{WHLo}) satisfy that $\{ M > 0,  \quad \Lambda < 0,  \quad M^{2}+  \Lambda J^{2} > 0, \quad q_{\beta}^{2} -  J^{2}q_{\alpha}^{2} < 0 \}$ which is compatible with (\ref{WormHolecondition}).
 Hence, according to the arguments given in (\ref{WHfNLED}), the solution (\ref{WHLo}) corresponds to a  five-parametric family of stationary and circularly symmetric (2+1)-dimensional traversable wormhole metric\footnote{ If the transition of the cosmological constant was;
$\Lambda: \Lambda_{0} \rightarrow \Lambda_{0} - \delta^{2}$, then $M^{2}+ \Lambda J^{2} =M^{2}+  (-\frac{M^{2}}{ J^{2}} - \delta^{2}) J^{2}= -\delta^{2} J^{2} < 0$, and therefore (\ref{WHLo}) would not be neither a black hole nor a wormhole's metric, and then there would not happen the transition black hole - wormhole.}.
 
 In summary, both (\ref{bHLo}) and (\ref{WHLo})  are solutions of general relativity  coupled to nonlinear electrodynamics with characteristic function $L(F) = \sqrt{-s F}$ and with electromagnetic field; $\boldsymbol{f} = f_{t\phi} dt \wedge d\phi $.  Despite the fact that  the cosmological constants of the solutions (\ref{bHLo}) and (\ref{WHLo}) differ very little one from another,  the  consequence is that  (\ref{bHLo}) and (\ref{WHLo}) describe different spacetimes.  The solution (\ref{bHLo}) describes a stationary and circularly symmetric (2+1)-dimensional  extreme black hole, whereas  (\ref{WHLo}) describes a stationary and circularly symmetric (2+1)-dimensional traversable wormhole. A feasible way to connect both spacetimes could be described by something like a VCC model.

%%%%%%%%%%%%%%%%%%%%%%%%%%%%%%%%%%%%%%%%%%%%%%%

\subsection{ Null cosmological constant ($\Lambda = 0$)  limit }\label{limL0}
 
Previously, in subsections \ref{btzBH} and \ref{newextremeBH},  it was analyzed the fundamental role that the cosmological constant plays for  interpreting  (\ref{NewSolution}) as a (2+1)-black hole; whereas in  subsections (\ref{WHfNLED}) and (\ref{transitionsBHWH}) the effects of NLED and a negative cosmological constant were presented. In this subsection  by  switching off $\Lambda$ we will examine the effect on the curvature of the spacetime, due only to the NLED.

{ \it\textbf{ Limit $\Lambda = 0$ } }\\

For $\Lambda = 0$ the solution (\ref{NewSolution}) becomes, 

\begin{equation}\label{newBHL0}
ds^{2} = - \left[ q_{\alpha} \left( - M +  \frac{J^{2}}{2r^{2}} \right)r +  q_{\beta} \sqrt{- M +  \frac{J^{2}}{4r^{2}}  }  \right]^{2} dt^{2}+ \frac{dr^{2}}{- M+ \frac{J^{2}}{4r^{2}}} 
+  r^{2}\left[d\phi - J \left(\frac{ q_{\beta} }{2 r^{2}}+ \frac{ q_{\alpha} }{r}\sqrt{- M + \frac{J^{2}}{4r^{2}}} \right) dt \right]^{2}.  
\end{equation}
In this case the zeroes of the functions $N^{2}(r)$ and $f^{2}(r)$  are,

\begin{eqnarray}
&& x_{_{1 ,N}}= \sqrt{ \frac{ J^{2}q_{\alpha}^{2}- q_{\beta}^{2}+  q_{\beta}\sqrt{ q_{\beta}^{2}- J^{2}q_{\alpha}^{2} } }{ 2 
M q_{\alpha}^{2} }}, \quad 
x_{_{2,N}}= \sqrt{ \frac{ J^{2}q_{\alpha}^{2}- q_{\beta}^{2} - q_{\beta}\sqrt{ q_{\beta}^{2}- J^{2}q_{\alpha}^{2} } }{ 2 
M q_{\alpha}^{2} }} , \label{x1x2NL0}\\
&&x_{_{1,f}} = \frac{J}{2\sqrt{M}}, \quad
x_{_{2,f}} =-\frac{J}{2\sqrt{M}}. \label{x1x2fL0}
\end{eqnarray}

 In contrast to the conditions imposed to the parameters in \ref{newextremeBH}, so that the solution represented a black hole, in the present case it will not be possible to restrict the parameters ($M$, $J$, $q_{\alpha}$, $q_{\beta}$), in order that the solution (\ref{newBHL0}) has event horizons.  However the parameters  can be adjusted,  as will be described next, in order to the solution (\ref{newBHL0})  corresponds to a wormhole structure given by (\ref{whAnsatz}), i.e.,  adjusting  $N^{2}(r) = e^{2\Phi(r)}$ and $f^{2}(r) = 1 - \frac{b(r)}{r}$ so that there exist  a wormhole throat  $r_{0}$, such that    $b(r_{0}) = r_{0}$, and that $e^{2\Phi(r)}$ be  a well defined and  finite function, for all $r\in [r_{0}, \infty)$. 

In this context, considering  the zeroes of $f^{2}(r) = 1-\frac{b(r)}{r} = - M+ \frac{J^{2}}{4r^{2}}$, we will assume that the WH throat, 
$r_{0}$, is localized at,

\begin{equation}
r_{0} = \frac{|J|}{2\sqrt{M}} 
\end{equation}
However, according to $e^{\Phi(r)} = q_{\alpha} r \left( - M +  \frac{J^{2}}{2r^{2}} \right) +  q_{\beta} \sqrt{- M +  \frac{J^{2}}{4r^{2}}  }  $,  it turns out that for $r\in(r_{0},\infty)$ the square root  $\sqrt{ - M +  \frac{J^{2}}{4r^{2}} } \not\in \mathbb{R}$, therefore the red-shift function is not defined for $r\in(r_{0},\infty)$. One simple way to avoid this problem would be switching off $q_{\beta}$. Doing so  the metric (\ref{newBHL0}) becomes, 

\begin{equation}\label{newBHL0qb}
ds^{2} = - q^{2}_{\alpha} \left( - M +  \frac{J^{2}}{2r^{2}} \right)^{2} r^{2} dt^{2}+ \frac{dr^{2}}{- M+ \frac{J^{2}}{4r^{2}}} 
+  r^{2}\left[d\phi - J \left(\frac{ q_{\alpha} }{r}\sqrt{- M + \frac{J^{2}}{4r^{2}}} \right) dt \right]^{2},  
\end{equation}
and in this case the red-shift function will have the appropriate behavior, compatible with that of a wormhole solution. i.e., $e^{2\Phi(r)} = q^{2}_{\alpha} r^{2} \left( - M +  \frac{J^{2}}{2r^{2}} \right)^{2} > 0,$ for $r\in(r_{0},\infty)$. 

 But recall that we have to check the flaring out condition still: it turns out that for this configuration it is not fulfilled $1-\frac{b(r)}{r}\geq0$,  $\forall r \in [r_{0},\infty)$. In fact, in this case what is found is that,

\begin{equation}\label{nonTWH}
1-\frac{b(r)}{r} = - M + \frac{J^{2}}{4r^{2}} \leq 0 ,\quad \forall r \in [r_{0},\infty)
\end{equation}
Then  it is easy to see that in this case the metric (\ref{newBHL0qb}) does not satisfy the flaring  out condition. Briefly, given that the shape function is $b(r) = ( 1 + M - \frac{J^{2}}{4r^{2}})r$, one finds that, $b(r) - rb_{,r}(r) = - \frac{J^{2}}{2r}$. Thus, one may conclude that,

\begin{equation}\label{nonTWHFC}
b(r) - rb_{,r}(r) < 0  \quad \forall r \in [r_{0},\infty).
\end{equation}
Hence the flaring out condition does not hold, and for that reason the solution (\ref{newBHL0qb}) does not correspond to  a traversable wormhole. This contrasts with what happens, in \ref{WHfNLED}, with (\ref{Shape}) and (\ref{flaringcondition}),  case in which due to the influence (or contribution) of a negative cosmological constant, $\Lambda$, it is fulfilled that; $1-\frac{b(r)}{r} \geq0$, together with the flaring out condition $b(r) - rb_{,r}(r)>0$, in the whole range of $r \in [r_0, \infty).$

\section{Conclusions}\label{Concl}

In this work we have determined the general form of the electromagnetic field tensor $f_{ab}$ in nonlinear electrodynamics with Lagrangian $L(F)$, for stationary cyclic (2+1) spacetimes. The expression of $f_{ab}$ depends on three parameters, $a, b$ and $c$, and depending if $c \ne 0$ or $c=0$, two disjoint branches of electromagnetism in (2+1) are defined. This result generalizes to NLED the corresponding Theorem for Maxwell electrodynamics, proved by Ay\'on, Cataldo and Garc\'ia, in Ref. \cite{Ayon}.

In the second part of the paper, for the $c\ne 0$ branch we have posed the Einstein-anti-de Sitter-NLED equations and then, choosing a particular NLED,  $L= \sqrt{-sF}$ (Einstein-power-Maxwell electrodynamics), we have determined a five-parametric stationary cyclic (2+1) family of solutions. The parameters are: mass $M$, angular momentum $J$, cosmological constant or de Sitter parameter $\Lambda$ and two electromagnetic parameters, $q_{\alpha}$ and $q_{\beta}$. The interpretation of the parameters was guided by the corresponding forms of the limiting solutions, being the uncharged limit the BTZ black hole. The derived family can be adscribed the interpretation of a wormhole, as it fulfills the conditions of a traversable wormhole, namely, the existence of a throat, the absence of horizons  as well as the violation of NEC.
For a particular value of the anti-de Sitter parameter, $\Lambda_0=-M^2/J^2$  the metric becomes the one of a black hole with one horizon and zero surface gravity, representing then an extreme black hole in anti-de Sitter space.

When the value of $\Lambda_0=-M^2/J^2$ is slightly increased, the black hole structure is lost and  a wormhole spacetime arises.
The thermodynamics of these metrics, that depending on the values of the parameters may be interpreted as  wormhole or black hole  is worth to be studied  as well as it looks promising for finding phase transitions and other interesting effects.

For the NLED with $L(F)=\sqrt{-sF}$ and in the branch $c\neq0$, the limit of vanishing anti-de Sitter parameter was analyzed. We showed that in this case the parameters of the solution cannot be restricted in order that the metric be interpreted as a black hole, or a traversable wormhole solution. For that reason, and the analysis developed in this work (where we were able to build solutions of type BH and WH), one can conclude  that for the derived family of solutions, the construction of  black holes or wormholes  will only be feasible by including the contribution to the curvature of spacetime coming from a negative cosmological constant.
 
The quantum properties of the derived family of solutions can also be of interest  in a similar way than the ones  characterizing the BTZ black hole, and in fact the metric functions are polynomials slightly more complicated that the ones characterizing the BTZ solution.

\section*{Acknowledgements}
%%%%%

Partial financial support by CONACYT through project No. 284489 is acknowledged.
 
%%%%%
%%%%%%%%%%%%%%%%%%%%%%%%%%%%%%%%%%%%%%%%%%%%%%%%%%%%%%%%%%%%%%%%

\end{document}